# Magnetic properties of artificially prepared highly ordered two-dimensional shunted and unshunted Nb–AlO$_X$–Nb Josephson junctions arrays

Fernando M. Araújo-Moreira
*Department of Physics and Physical Engineering - UFSCar,*
*Laboratory of Materials and Devices,*
*Multidisciplinary Center for the Development of Ceramic Materials,*
*Caixa Postal 676, São Carlos/SP 13565-905, BRAZIL*

Sergei A. Sergeenkov
*Bogoliubov Laboratory of Theoretical Physics, Joint Institute for Nuclear Research,*
*141980 Dubna, Moscow Region, Russia*

## Abstract

Josephson junction arrays (JJA) have been actively studied for decades. However, they continue to contribute to a wide variety of intriguing and peculiar phenomena. To name just a few recent examples, it suffice to mention the so-called paramagnetic Meissner effect and related reentrant temperature behavior of AC susceptibility, observed both in artificially prepared JJA and granular superconductors. Employing mutual-inductance measurements and using a high-sensitive home-made bridge, we have thoroughly investigated the temperature and magnetic field dependence of complex AC susceptibility of artificially prepared highly ordered (periodic) two-dimensional Josephson junction arrays (2D-JJA) of both shunted and unshunted Nb–AlO$_X$–Nb tunnel junctions

In this Chapter, we report on three phenomena related to the magnetic properties of 2D-JJA: (a) the influence of non-uniform critical current density profile on magnetic field behavior of AC susceptibility; (b) the origin of dynamic reentrance and the role of the Stewart-McCumber parameter, $\beta_C$, in the observability of this phenomenon, and (c) the manifestation of novel geometric effects in temperature behavior of AC magnetic response. Firstly, we present evidences for the existence of local type non-uniformity in the periodic (globally uniform) unshunted 2D-JJA. Specifically, we found that in the mixed state region AC susceptibility $\chi(T, h_{AC})$ can be rather well fitted by a single-plaquette approximation of the overdamped 2D-JJA model assuming a non-uniform distribution of the critical current density within a single junction. According to the current paradigm, paramagnetic Meissner effect (PME) can be related to the presence of $\pi$-junctions, either resulting from the presence of magnetic impurities in the junction or from unconventional pairing symmetry. Other possible explanations of this phenomenon are based on flux trapping and flux compression effects including also an important role of the surface of the sample. Besides, in the experiments with unshunted 2D-JJA, we have previously reported that PME manifests itself through a dynamic reentrance (DR) of the AC magnetic susceptibility as a function of temperature. Using an analytical expression we successfully fit our experimental data and demonstrate that the dynamic reentrance of AC susceptibility is directly linked to the value of $\beta_C$. By simultaneously varying the parameter $\beta_L$, a phase diagram $\beta_C$-$\beta_L$ is plotted which demarcates the border between the reentrant and non-reentrant behavior. We show that only arrays with sufficiently large value of $\beta_C$ will exhibit the dynamic reentrance behavior and hence PME. The last topic reviewed in this Chapter is related to the step-like structure observed when the resolution of home-made mutual-inductance bridge is improved. That structure (with the number of steps $n = 4$ for all AC fields) has been observed in the temperature dependence of AC susceptibility in unshunted 2D-JJA with $\beta_L(4.2K) = 30$. We were able to successfully fit our data assuming that steps are related to the geometric properties of the plaquette. The number of steps $n$ corresponds to the number of flux quanta that can be screened by the maximum critical current of the junctions. The steps are predicted to manifest themselves in arrays with the inductance related parameter $\beta_L(T)$ matching a "*quantization*" condition $\beta_L(0)=2\pi(n+1)$.

# I. Introduction

Artificially prepared two-dimensional Josephson junctions arrays (2D-JJA) consist of highly ordered superconducting islands arranged on a symmetrical lattice coupled by Josephson junctions (Fig. **1**), where it is possible to introduce a controlled degree of disorder. In this case, a 2D-JJA can be considered as the limiting case of an extreme inhomogeneous type-II superconductor, allowing its study in samples where the disorder is nearly exactly known. Since 2D-JJA are artificial, they can be very well characterized. Their discrete nature, together with the very well-known physics of the Josephson junctions, allows the numerical simulation of their behavior (see very interesting reviews by Newrock *et al.*[1] and by Martinoli *et al.*[2] on the physical properties of 2D-JJA).

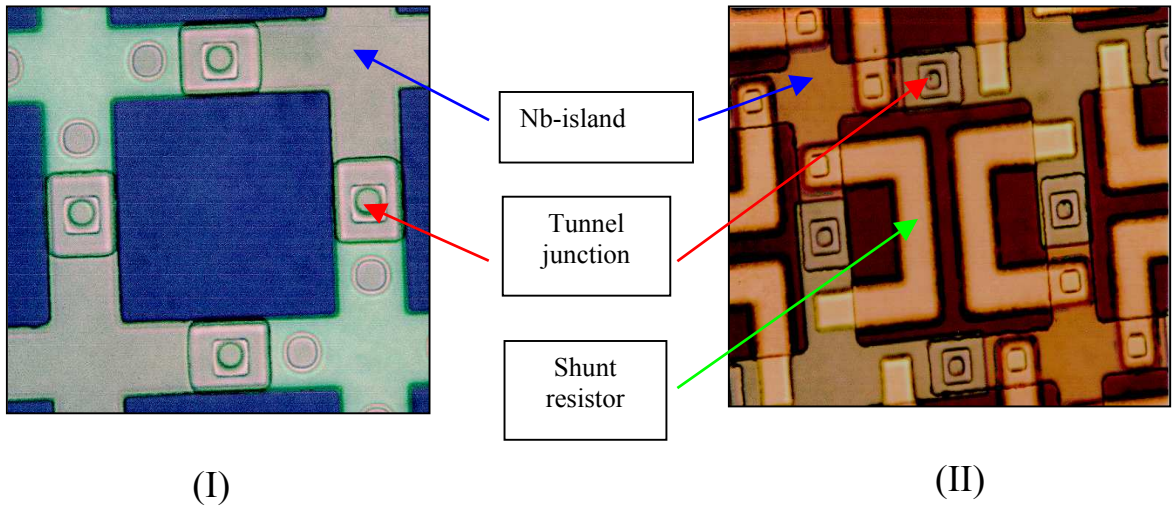

**Figure 1** – Photograph of unshunted (I) and shunted (II) Josephson junction arrays.

Many authors have used a parallelism between the magnetic properties of 2D-JJA and granular high-temperature superconductors (HTS) to study some controversial features of HTS. It has been shown that granular superconductors can be considered as a collection of superconducting grains embedded in a weakly superconducting - or even normal - matrix. For this reason, granularity is a term specially related to HTS, where magnetic and transport properties of these materials are usually manifested by a two-component response. In this scenario, the first component represents the *intragranular* contribution, associated to the grains exhibiting ordinary superconducting properties, and the second one, which is originated from *intergranular* material, is associated to the weak-link structure, thus, to the Josephson junctions network[3-6]. For single-crystals and other nearly-perfect structures, granularity is a more subtle feature that can be envisaged as the result of a symmetry breaking. Thus, one might have granularity on the nanometric scale, generated by localized defects like impurities, oxygen deficiency, vacancies, atomic substitutions and the genuinely *intrinsic* granularity associated with the layered structure of perovskites. On the micrometric scale, granularity results from the existence of extended defects, such as grain and twin boundaries. From this picture, granularity could have many contributions, each one with a different volume fraction [7-10]. The small coherence length of HTS implies that any imperfection may contribute to both the weak-link properties and the flux pinning. This leads to many interesting peculiarities and anomalies, many of which have been tentatively explained over the years in terms of the granular character of HTS materials.

One of the controversial features of HTS elucidated by studying the magnetic properties of 2D-JJA is the so-called Paramagnetic Meissner Effect (PME), also known as Wohlleben Effect. In this case, one considers first the magnetic response of a granular superconductor submitted to either an AC or DC field of small magnitude. This field should be weak enough to guarantee that the critical current of the intergranular material is not exceeded at low temperatures. After a zero-field cooling (ZFC) process which consists in cooling the sample from above its critical temperature ($T_C$) with no applied magnetic field, the magnetic response to the application of a magnetic field is that of a perfect diamagnet. In this case, the intragranular screening currents prevent the magnetic field from entering the grains, whereas intergranular currents flow across the sample to ensure a null magnetic flux throughout the whole specimen. This temperature dependence of the magnetic response gives rise to the well-known double-plateau behavior of the DC susceptibility and the corresponding double-drop/double-peak of the complex AC magnetic susceptibility[7-11]. On the other hand, by cooling the sample in the presence of a magnetic field, by following a field-cooling (FC) process, the screening currents are restricted to the intragranular contribution (a situation that remains until the temperature reaches a specific value below which the critical current associated to the intragrain component is no longer equal to zero). It has been experimentally confirmed that intergranular currents may contribute to a magnetic behavior that can be either paramagnetic or diamagnetic. Specifically, where the intergranular magnetic behavior is paramagnetic, the resulting magnetic susceptibility shows a striking reentrant behavior. All these possibilities about the signal and magnitude of the magnetic susceptibility have been extensively reported in the literature, involving both LTS and HTS materials [12-15]. The reentrant behavior mentioned before is one of the typical signatures of PME. We have reported its occurrence as a reentrance in the temperature behavior of the AC magnetic susceptibility of 2D-JJA [16,17]. Thus, by studying 2D-JJA, we were able to demonstrate that the appearance of PME is simply related to trapped flux and has nothing to do with manifestation of any sophisticated mechanisms, like the presence of pi-junctions or unconventional pairing symmetry.

In this Chapter we report on three phenomena related to the magnetic properties of 2D-JJA: (a) the influence of non-uniform critical current density profile on magnetic field behavior of AC susceptibility; (b) the observability of dynamic reentrance and the role of the Stewart-McCumber parameter, $\beta_C$, in this phenomenon, and (c) the manifestation of novel geometric effects in temperature behavior of AC magnetic response. To perform this work, we have used numerical simulations and both the mutual-inductance and the scanning SQUID microscope experimental techniques.

The paper is organized as follows. In Sec. II we outline the main concepts related to the mutual-inductance technique (along with the physical meaning of the measured output voltage) as well as the scanning SQUID microscope experimental technique. In Sec. III we review the numerical simulations based on a unit cell containing four Josephson junctions. In Sec. IV we describe the influence of non-uniform critical current density profile on magnetic field behavior of AC susceptibility and discuss the obtained results. In Sec. V we study the origin of dynamic reentrance and discuss the role of the Stewart-McCumber parameter in the observability of this phenomenon. In Sec. VI we present the manifestation of completely novel geometric effects recently observed in the temperature behavior of AC magnetic response. And finally, in Sec. VII we summarize the main results of the present work.

## II. The mutual-inductance technique

Complex AC magnetic susceptibility is a powerful low-field technique to determine the magnetic response of many systems, like granular superconductors and Josephson junction arrays. It has been successfully used to measure several parameters such as critical temperature, critical current density and penetration depth in superconductors. To measure samples in the shape of thin films, the so-called *screening method* has been developed. It involves the use of primary and secondary coils, with diameters smaller than the dimension of the sample. When these coils are located near the surface of the film, the response, i.e., the complex output voltage $V$, does not depend on the radius of the film or its properties near the edges. In the reflection technique [18], an excitation coil (primary) coaxially surrounds a pair of counter-wound pick up coils (secondaries). When there is no sample in the system, the net output from these secondary coils is close to zero since the pick up coils are close to identical in shape but are wound in opposite directions. The sample is positioned as close as possible to the set of coils, to maximize the induced signal on the pick up coils (Figure 2).

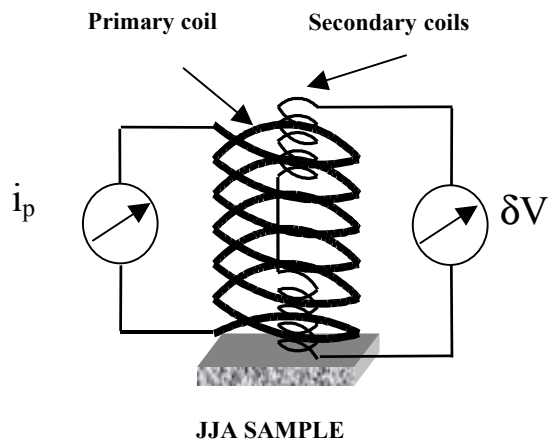

**Figure 2** – Screening method in the reflection technique, where an excitation coil (primary) coaxially surrounds a pair of counter-wound pick up coils (secondaries).

An alternate current sufficient to create a magnetic field of amplitude $h_{AC}$ and frequency $f$ is applied to the primary coil. The output voltage of the secondary coils, $V$, is a function of the complex susceptibility, $\chi_{AC} = \chi' + i\chi''$, and is measured through the usual lock-in technique. If we take the current on the primary as a reference, $V$ can be expressed by two orthogonal components. The first one is the inductive component, $V_L$ (in phase with the time-derivative of the reference current) and the second one the quadrature resistive component, $V_R$ (in phase with the reference current). This means that $V_L$ and $V_R$ are correlated with the average magnetic moment and the energy losses of the sample, respectively.

We used the screening method in the reflection configuration to measure $\chi_{AC}(T)$ of Josephson junction arrays. Measurements were performed as a function of the temperature T (1.5K < T < 15K), the amplitude of the excitation field $h_{AC}$ (1 mOe < $h_{AC}$ < 10 Oe), and the external magnetic field $H_{DC}$ (0 < $H_{DC}$ < 100 Oe) parallel with the plane of the sample (Figure 3).

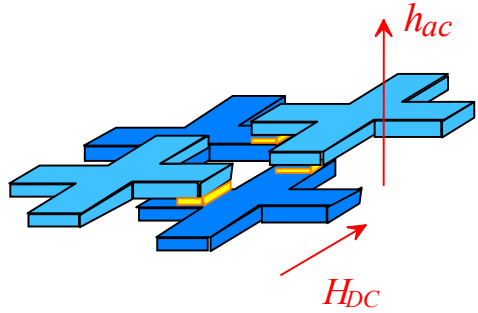

**Figure 3** - Sketch of the experimental setup, where the excitation field $h_{ac}$ and the external magnetic field $H_{dc}$ are respectively perpendicular and parallel to the plane of the sample.

The frequency in the experiments reported here was fixed at f = 1.0 kHz. The typical dimensions of the coils and samples are depicted in Fig. 4 The susceptometer was positioned inside a double wall μ-metal shield, screening the sample region from Earth's magnetic field.

For a complete description of this technique, let us study now the relation between the measured complex voltage, $V = V_L + iV_R$, and the components of the AC magnetic susceptibility, χ' and χ". We assume that the current in the drive coil (primary) is given by $I_D e^{i\omega t}$, which creates at the sample an average magnetic field $H_D e^{i\omega t}$. Considering the section of the sample as a simple loop, we model its response as an impedance $Z_S$ in series with a geometrical inductance, $L_g$. The impedance depends on the material parameters as well as the size of the loop. For a normal metal sample, $Z_S = 2\pi\rho(rt)\Delta r$, with $\rho$ the resistivity of the material, r the radius of the loop, t the thickness of the sample, and $\Delta r$ the width of the loop.

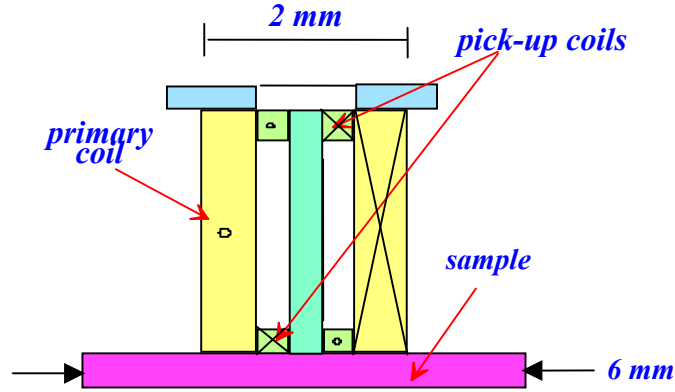

**Figure 4** - Typical dimensions of the coils and samples.

We can obtain equivalent equations for the specific case of a superconducting material. The equation relating the drive field to the current response $I_S$ of the loop is given by:

$$-\frac{\partial \Phi_{ext}}{\partial t} = -i\omega\mu_0 H_D e^{i\omega t} A = I_S(Z_S + i\omega L_g) \quad \text{(II.1)}$$

where A is the area of the loop. Taking $Z_S = X = iY$, Eq. (II.1) reduces to:

$$I_S = \frac{-iA\omega\mu_0 H_D e^{i\omega t}}{X + i(Y + \omega L_g)} \quad \text{(II.2)}$$

The induced voltage in the pick-up coil is given by:

$$-M_{SP}(i\omega I_S) = V_P \quad \text{(II.3)}$$

where $M_{SP}$ is the mutual inductance between the sample and the pickup coil. Combining Eqs. (II.2) and (II.3), we obtain:

$$V_P = -\frac{\omega^2 A M_{SP}\mu_0 H_D e^{i\omega t}}{X + i(Y + \omega L_g)} \quad \text{(II.4)}$$

To obtain the magnetic susceptibility, we first find a relationship between the effective magnetization <M> of the loop and $I_S$. Since $B = \mu_0(H + M)$, we may write:

$$\mu_0[<H> + <M>]A = <B>A = \Phi \quad \text{(II.5)}$$

From this, we identify the magnetic flux due to the current in the sample as being proportional to the average magnetization:

$$\mu_0 <M> A = L_g I_S \qquad (II.6)$$

Combining Eqs. (II.2) and (II.6), gives:

$$-\frac{i\omega L_g H_D e^{i\omega t}}{X + i(Y + \omega L_g)} = <M> = (\chi' - i\chi'') H_D e^{i\omega t} \qquad (II.7)$$

where we have neglected higher harmonics considering the response of the loop given by the average magnetization:

$$<M> = (\chi' - i\chi'') H_D e^{i\omega t} \qquad (II.8)$$

On the other hand, since the pickup coil is counter wound, it only responds to $dM/dt$, so that:

$$V_P \propto -\frac{\partial M}{\partial t} \propto (-\omega\chi'' - i\omega\chi') H_D e^{i\omega t} \qquad (II.9)$$

From Eqs. (II.2) and (II.6)- (II.8), we obtain:

$$\frac{\mu_0 M_{SP} A \omega}{L_g}(-\chi'' - i\chi') H_D e^{i\omega t} = V_P = V_P' + i V_P'' \qquad (II.10)$$

which agrees with Eq.(II.9). From Eq. (II.7) we can write:

$$\chi' = \frac{\omega L_g Y + \omega^2 L_g^2}{X^2 + (Y + \omega L_g)^2} \qquad (II.11a)$$

$$\chi'' = \frac{\omega L_g X}{X^2 + (Y + \omega L_g)^2} \qquad (II.11b)$$

To get the complete response of a real sample, these equations should be integrated over the whole specimen. For the special case of a superconducting loop far below $T_C$, where we can neglect the normal channel in a two-fluid model, the induced EMF in a magnetic field $H_D e^{i\omega t}$ is still given by $\varepsilon = -i\omega A \mu_0 H_D e^{i\omega t}$. The loop has now a kinetic inductance $L_K$ as well as a geometrical inductance $L_g$ so that the current is given by $-i\omega A \mu_0 H_D e^{i\omega t} = i\omega(L_K + L_g) I_S$, or $I_S = -(A\mu_0 H_D e^{i\omega t})/(L_K + L_g)$. Eq. (II.6) implies that the magnetization is:

$$<M> = -\frac{L_g}{L_K + L_g} H_D e^{i\omega t} \qquad (II.12)$$

or, alternatively, that:

$$\chi' = -\frac{L_g}{L_K + L_g} \quad (II.13)$$
$$\chi'' = 0$$

which agrees with Eqs. (II.11) setting $X = 0$ and $Y = \omega L_K$.

Therefore, we have:

$$\chi' \propto V_L \quad (II.14a)$$
$$\chi'' \propto V_R \quad (II.14b)$$

This means that by measuring the output voltage from the secondary coils, we can obtain the components of the complex AC magnetic susceptibility, $\chi$, as we stated in the beginning.

# III. Numerical simulations

We have found that all the experimental results obtained from the magnetic properties of 2D-JJA can be qualitatively explained by analyzing the dynamics of a single unit cell in the array [16, 17].

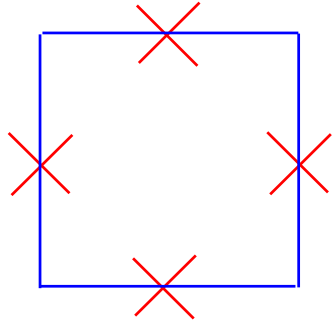

**Figure 5** – Unit cell of the array, containing a loop with four identical junctions.

In our experiments, the unit cell is a loop containing four junctions (Fig. 5) and the measurements correspond to ZFC AC magnetic susceptibility. We model a single unit cell as having four identical junctions, each with capacitance $C_J$, quasi-particle resistance $R_J$ and critical current $I_C$. We apply an external field of the form:

$$H_{ext} = h_{AC}\cos(\omega t) \quad \text{(III.1)}$$

The total magnetic flux, $\Phi_{TOT}$, threading the four-junction superconducting loop is given by:

$$\Phi_{TOT} = \Phi_{EXT} + LI \quad \text{(III.2)}$$

where $\Phi_{EXT} = \mu_0 a^2 H_{EXT}$ with $\mu_0$ being the vacuum permeability, I is the circulating current in the loop, L is the inductance of the loop and $\Phi_{EXT}$ is the flux related to the applied magnetic field. Therefore the total current is given by:

$$I = I_C \sin\gamma_i + \frac{\Phi_0}{2\pi R_J}\frac{d\gamma_i}{dt} + \frac{C_J\Phi_0}{2\pi}\frac{d^2\gamma_i}{dt^2} \quad \text{(III.3)}$$

Here, $\gamma_i$ is the superconducting phase difference across the $i$th junction and $I_C$ is the critical current of each junction. In the case of our model with four junctions, the fluxoid quantization condition, which relates each $\gamma_i$ to the external flux, is:

$$\gamma_i = \frac{\pi}{2}n - \frac{\pi}{2}\frac{\Phi_{TOT}}{\Phi_0} \qquad (III.4)$$

where *n* is an integer and, by symmetry, we assume:

$$\gamma_1 = \gamma_2 = \gamma_3 = \gamma_4 = \gamma_i \qquad (III.5)$$

In the case of an oscillatory external magnetic field of the form of Eq. (III.1), the magnetization is given by:

$$M = \frac{LI}{\mu_0 a^2} \qquad (III.6)$$

It may be expanded as a Fourier series in the form:

$$M(t) = h_{AC}\sum_{n=0}^{\infty}[\chi_n^{'}\cos(n\omega t) + \chi_n^{''}\sin(n\omega t)] \qquad (III.7)$$

We calculated $\chi'$ and $\chi''$ through this equation. Both Euler and fourth-order Runge-Kutta integration methods provided the same numerical results. In our model we do not include other effects (such as thermal activation) beyond the above equations. In this case, the temperature-dependent parameter is the critical current of the junctions, given to good approximation by [19]:

$$I_C(T) = I_C(0)\sqrt{1 - \frac{T}{T_C}}\tanh\left[1.54\frac{T_C}{T}\sqrt{1 - \frac{T}{T_C}}\right] \qquad (III.8)$$

We calculated $\chi_1$ as a function of T. $\chi_1$ depends on the parameter $\beta_L$, which is proportional to the number of flux quanta that can be screened by the maximum critical current in the junctions, and the parameter $\beta_C$, which is proportional to the capacitance of the junction:

$$\beta_L(T) = \frac{2\pi L I_C(T)}{\Phi_0} \qquad (III.9)$$

$$\beta_C(T) = \frac{2\pi I_C C_J R_J^2}{\Phi_0} \qquad (III.10)$$

# IV. Influence of non-uniform critical current density profile on magnetic field behavior of AC susceptibility

Despite the fact that Josephson junction arrays (2D-JJA) have been actively studied for decades, they continue to contribute to the variety of intriguing and peculiar phenomena. To name just a few recent examples, it suffice to mention the so-called paramagnetic Meissner effect and related reentrant temperature behavior of AC susceptibility, observed both in artificially prepared 2D-JJA and granular superconductors (for recent reviews on the subject matter, see Refs. [20–24] and further references therein). So far, most of the investigations have been done assuming an ideal (uniform) type of array. However, it is quite clear that, depending on the particular technology used for preparation of the array, any real array will inevitably possess some kind of non-uniformity, either global (related to a random distribution of junctions within array) or local (related to inhomogeneous distribution of critical current densities within junctions). For instance, recently a comparative study of the magnetic remanence exhibited by disordered (globally non-uniform) 3D-JJA in response to an excitation with an AC magnetic field was presented[25]. The observed temperature behavior of the remanence curves for arrays fabricated from three different materials (Nb, $YBa_2Cu_3O_7$ and $La_{1.85}Sr_{0.15}CuO_4$) was found to follow the same universal law regardless of the origin of the superconducting electrodes of the junctions which form the array. In the section, through an experimental study of complex AC magnetic susceptibility $\chi(T,h_{ac})$ of the periodic (globally uniform) 2D-JJA of unshunted Nb–AlOx–Nb junctions, we present evidence for existence of the local type non-uniformity in our arrays. Here, $h_{AC}$ corresponds to the amplitude of excitation field. Specifically, we found that in the mixed state region $\chi(T,h_{ac})$ can be rather well fitted by a single-plaquette approximation of the over-damped 2D-JJA model assuming a non-uniform (Lorentz-like) distribution of the critical current density within a single junction.

Our samples consisted of $100 \times 150$ unshunted tunnel junctions. The unit cell had square geometry with lattice spacing a = 46 μm and a junction area of $5 \times 5$ μm$^2$. The critical current density for the junctions forming the arrays was about 600 A/cm$^2$ at 4.2 K, giving thus $I_C$ = 150 μA for each junction. We used the screening method[26] in the reflection configuration to measure the complex AC susceptibility $\chi = \chi'+i\chi''$ of our 2D-JJA (for more details on the experimental technique and set-ups see [27–29]). Fig. 6 shows the obtained experimental data for the complex AC susceptibility $\chi(T,h_{ac})$ as a function of $h_{ac}$ for a fixed temperature below $T_C$. As is seen, below 50 mOe (which corresponds to a Meissner-like regime with no regular flux present in the array) the susceptibility, as expected, practically does not depend on the applied magnetic field, while in the mixed state (above 50 mOe) both $\chi'(T,h_{ac})$ and $\chi''(T,h_{ac})$ follow a quasi-exponential field behavior of the single junction Josephson supercurrent (see below).

To understand the observed behavior of the AC susceptibility, in principle one would need to analyze the flux dynamics in our over-damped, unshunted 2D-JJA. However, given a well-defined (globally uniform) periodic structure of the array, to achieve our goal it is sufficient to study just a single unit cell (plaquette) of the array. (It is worth noting that the single-plaquette approximation proved successful in treating the temperature reentrance phenomena of AC susceptibility in ordered 2D-JJA[24,27,28] as well as magnetic remanence in disordered 3D-JJA[25]). The unit cell is a loop containing four identical Josephson junctions. Since the inductance of each loop is $L = \mu_0 a = 64$ pH and the critical current of each junction is $I_C = 150$ µA, for the mixed-state region (above 50 mOe) we can safely neglect the self-field effects because in this region the inductance related flux $\Phi_L(t) = LI(t)$ (here I(t) is the total current circulating in a single loop[29]) is always smaller than the external field induced flux $\Phi_{ext}(t) = B_{ac}(t) \cdot S$ (here $S \approx a^2$ is the projected area of a single loop, and $B_{ac}(t) = \mu_0 h_{ac} \cos(\omega t)$ is an applied AC magnetic field). Besides, since the length L and the width w of each junction in our array is smaller than the Josephson penetration depth, then:

$$\lambda_j = \sqrt{\frac{\Phi_0}{2\pi\mu_0 d j_{c0}}}$$

(where $j_{c0}$ is the critical current density of the junction, $\Phi_0$ is the magnetic flux quantum, and $d = 2\lambda_L + \xi$ is the size of the contact area with $\lambda_L(T)$ being the London penetration depth of the junction and $\xi$ an insulator thickness), namely $L \approx w \approx 5$ µm and $\lambda_j \approx 20$ µm (using $j_{c0} = 600$ A/cm$^2$ and $\lambda_L = 39$ nm for Nb at T = 4.2 K), we can adopt the small junction approximation[29] for the gauge-invariant superconducting phase difference across the i*th* junction (by symmetry we assume that[27,28] $\phi_1 = \phi_2 = \phi_3 = \phi_4 = \phi_i$), then:

$$\phi_i(x,t) = \phi_0 + \frac{2\pi B_{ac}(t)d}{\phi_0} \cdot x \qquad (IV.1)$$

where $\phi_0$ is the initial phase difference. The net magnetization of the plaquette is $M(t) = SI_S(t)$, where the maximum upper current (corresponding to $\phi_0 = \pi/2$) through an inhomogeneous Josephson contact reads:

$$I_S(t) = \int_0^L dx \int_0^W dy\, j_c(x,y) \cos\phi_i(x,t) \qquad (IV.2)$$

For the explicit temperature dependence of the Josephson critical current density:

$$j_{c0}(T) = j_{c0}(0) \left[\frac{\Delta(T)}{\Delta(0)}\right] \tanh\left[\frac{\Delta(T)}{2k_B T}\right] \qquad (IV.3)$$

we used the well-known[30] analytical approximation for the BCS gap parameter (valid for all temperatures):

$$\Delta(T) = \Delta(0) \tanh\left(2.2\sqrt{\frac{T_C - T}{T}}\right)$$

where $\Delta(0) = 1.76 k_B T_C$.

In general, the values of $\chi'(T,h_{AC})$ and $\chi''(T,h_{AC})$ of the complex harmonic susceptibility are defined via the time dependent magnetization of the plaquette as follows:

$$\chi'(T,h_{ac}) = \frac{1}{\pi h_{AC}} \int_0^{2\pi} d(\omega t) \cos(\omega t) M(t) \qquad (IV.4)$$

$$\chi''(T,h_{AC}) = \frac{1}{\pi h_{AC}} \int_0^{2\pi} d(\omega t) \sin(\omega t) M(t) \qquad (IV.5)$$

Using Eqs. (IV.1)–(IV.5) to simulate the magnetic field behavior of the observed AC susceptibility of the array, we found that the best fit through all the data points and for all temperatures is produced assuming the following non-uniform distribution of the critical current density within a single junction[29]:

$$j_c(x,y) = j_{c0}(T)\left(\frac{L^2}{x^2+L^2}\right)\left(\frac{w^2}{y^2+w^2}\right) \qquad (IV.6)$$

It is worthwhile to mention that in view of Eq. (IV.2), in the mixed-state region the above distribution leads to approximately exponential field dependence of the maximum supercurrent $I_S(T,h_{AC}) \approx I_S(T,0)\exp(-h_{AC}/h_0)$ which is often used to describe critical-state behavior in type-II superconductors[31]. Given the temperature dependencies of the London penetration depth $\lambda_L(T)$ and the Josephson critical current density $j_{c0}(T)$, we find that:

$$h_0(T) = \frac{\Phi_0}{2\pi\mu_0 \lambda_j(T) L} \approx h_0(0)\cdot\left(\frac{T_C - T}{T_C}\right)^{1/4} \qquad (IV.7)$$

for the temperature dependence of the characteristic field near $T_C$. This explains the improvement of our fits (shown by solid lines in Fig. 6) for high temperatures because with increasing the temperature the total flux distribution within a single junction becomes more regular which in turn validates the use of the small-junction approximation.

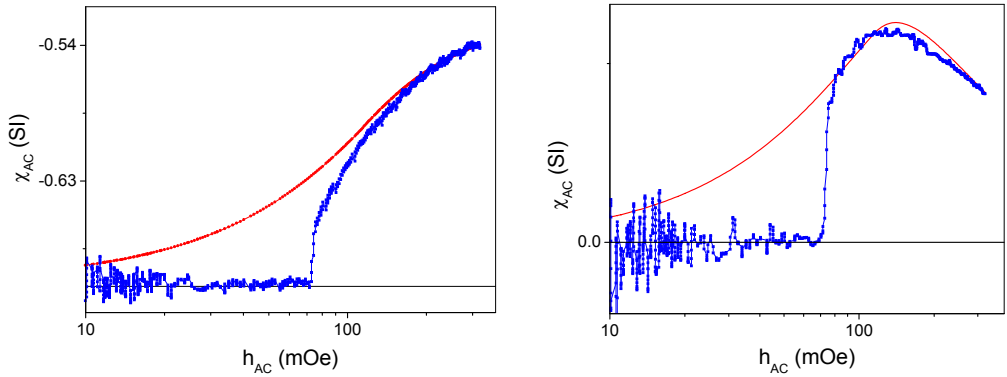

**(a)**

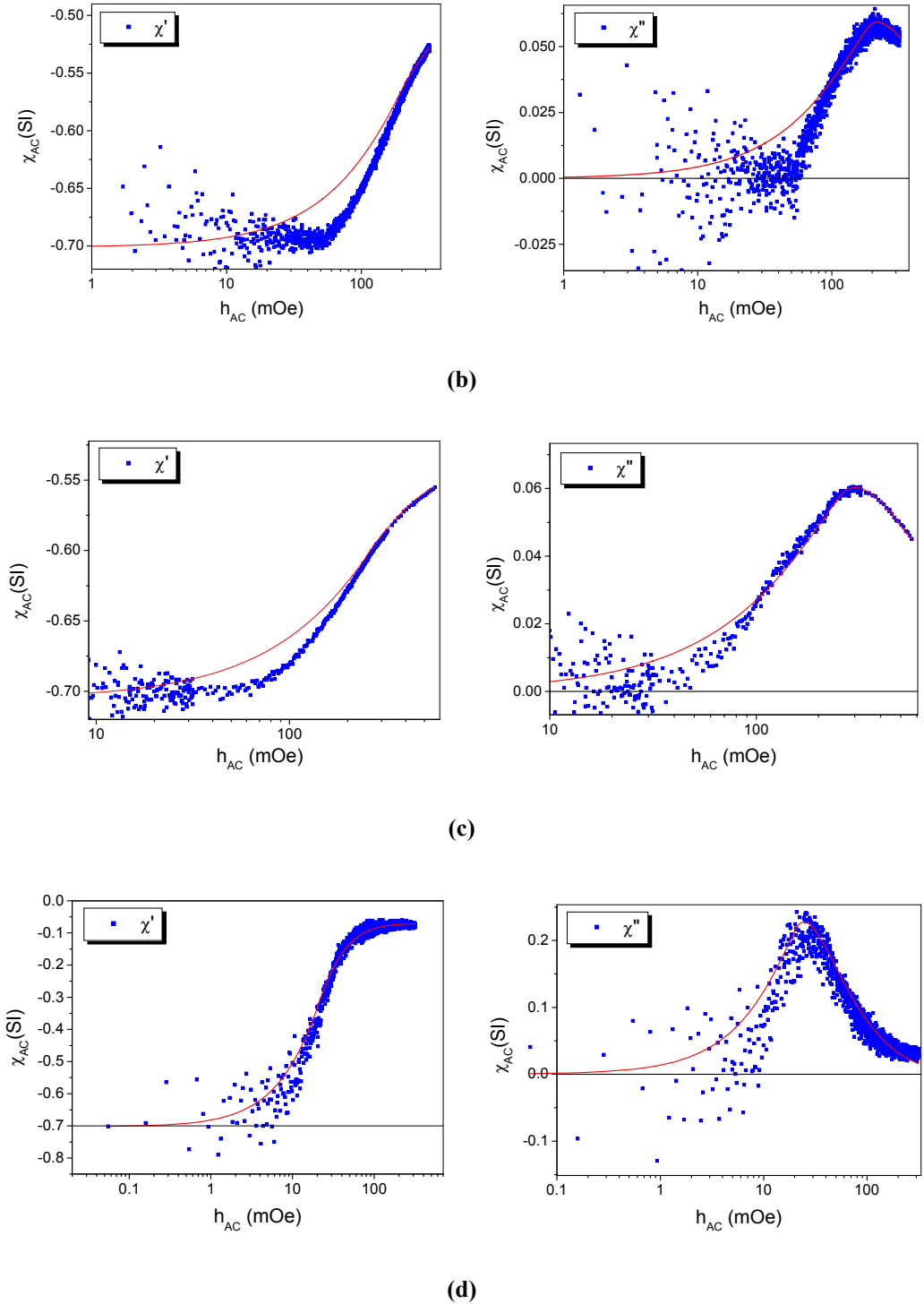

**Figure 6** – The dependence of both components of the complex AC magnetic susceptibilities, on AC magnetic field amplitude $h_{AC}$ for different temperatures: (a) T= 4.2 K, (b) T = 6 K, (c) T = 7.5 K, and (d) T = 8 K. Solid lines correspond to the fitting of the 2D-JJA model with non-uniform critical current profile for a single junction (see the text).

# V. On the origin of dynamic reentrance and the role of the Stewart-McCumber parameter

According to the current paradigm, paramagnetic Meissner effect (PME)[32-37], can be related to the presence of $\pi$-junctions[38], either resulting from the presence of magnetic impurities in the junction[39,40] or from unconventional pairing symmetry[41]. Other possible explanations of this phenomenon are based on flux trapping[42] and flux compression effects[43] including also an important role of the surface of the sample[34]. Besides, in the experiments with unshunted 2D-JJA, we have previously reported[44] that PME manifests itself through a dynamic reentrance (DR) of the AC magnetic susceptibility as a function of temperature. These results have been further corroborated by Nielsen *et al.*[45] and De Leo *et al.*[46] who argued that PME can be simply related to magnetic screening in multiply connected superconductors. So, the main question is: which parameters are directly responsible for the presence (or absence) of DR in artificially prepared arrays?

Previously (also within the single plaquette approximation), Barbara *et al.*[44] have briefly discussed the effects of varying $\beta_L$ on the observed dynamic reentrance with the main emphasis on the behavior of 2D-JJA samples with high (and fixed) values of $\beta_C$. However, to our knowledge, up to date no systematic study (either experimental or theoretical) has been done on how the $\beta_C$ value itself affects the reentrance behavior. In the present section of this review, by a comparative study of the magnetic properties of shunted and unshunted 2D-JJA, we propose an answer to this open question. Namely, by using experimental and theoretical results, we will demonstrate that only arrays with sufficiently large value of the Stewart-McCumber parameter $\beta_C$ will exhibit the dynamic reentrance behavior (and hence PME).

To measure the complex AC susceptibility in our arrays we used a high-sensitive home-made susceptometer based on the so-called screening method in the reflection configuration[47-49], as shown in previous sections. The experimental system was calibrated by using a high-quality niobium thin film.

To experimentally investigate the origin of the reentrance, we have measured $\chi'(T)$ for three sets of shunted and unshunted samples obtained from different makers (Westinghouse and Hypress) under the same conditions of the amplitude of the excitation field $h_{ac}$ (1 mOe < $h_{ac}$ <10 Oe), external magnetic field $H_{dc}$ (0 < $H_{dc}$ < 500 Oe) parallel to the plane of the sample, and frequency of AC field $\omega = 2\pi f$ (fixed at f = 20 kHz). Unshunted 2D-JJAs are formed by loops of niobium islands linked through Nb-AlO$_x$-Nb Josephson junctions while shunted 2D-JJAs have a molybdenum shunt resistor (with $R_{sh} \approx 2.2\Omega$) short-circuiting each junction (see Fig. 1). Both shunted and unshunted samples have rectangular geometry and consist of $100 \times 150$ tunnel junctions. The unit cell for both types of arrays has square geometry with lattice spacing a ≈ 46μm and a single junction area of $5 \times 5\mu m^2$. The critical current density for the junctions forming the arrays is about 600A/cm$^2$ at 4.2 K. Besides, for the unshunted samples $\beta_C(4.2K) \approx 30$ and $\beta_L(4.2K) \approx 30$, while for shunted samples $\beta_C(4.2K) \approx 1$ and $\beta_L(4.2K) \approx 30$ where $\beta_L$ and $\beta_C$ are given by expressions (III.9) and (III.10), respectively[50]. There, $C_j \approx 0.58pF$ is the capacitance, $R_j \approx 10.4\Omega$ the quasi-particle resistance (of unshunted array), and $I_C(4.2K) \approx 150\mu A$ the critical current of the Josephson junction. $\Phi_0$ is the quantum of magnetic flux. The parameter $\beta_L$ is proportional to the number of flux quanta that can be screened by the maximum critical current in the junctions, while the Stewart-McCumber parameter $\beta_C$ basically reflects the quality of the junctions in arrays.

It is well established that both magnetic and transport properties of any superconducting material can be described via a two-component response[51], the *intragranular* (associated with the grains exhibiting bulk superconducting properties) and *intergranular* (associated with weak-link structure) contributions[52,53]. Likewise, artificially prepared JJAs (consisting of superconducting islands, arranged in a symmetrical periodic lattice and coupled by Josephson junctions) will produce a similar response[54].

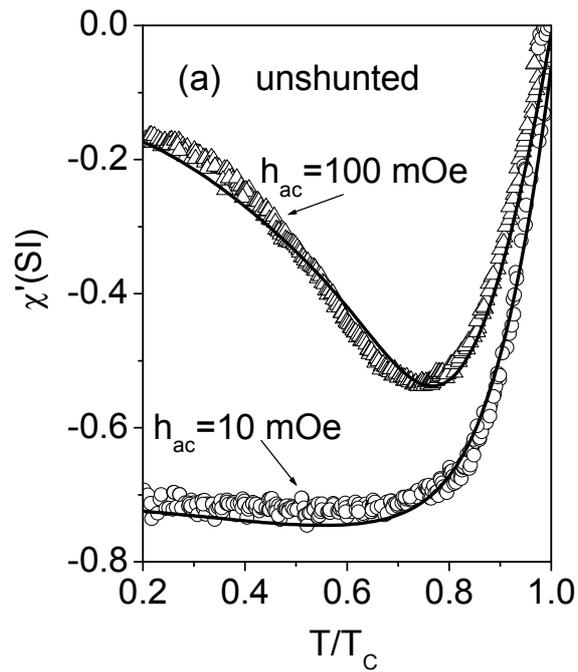

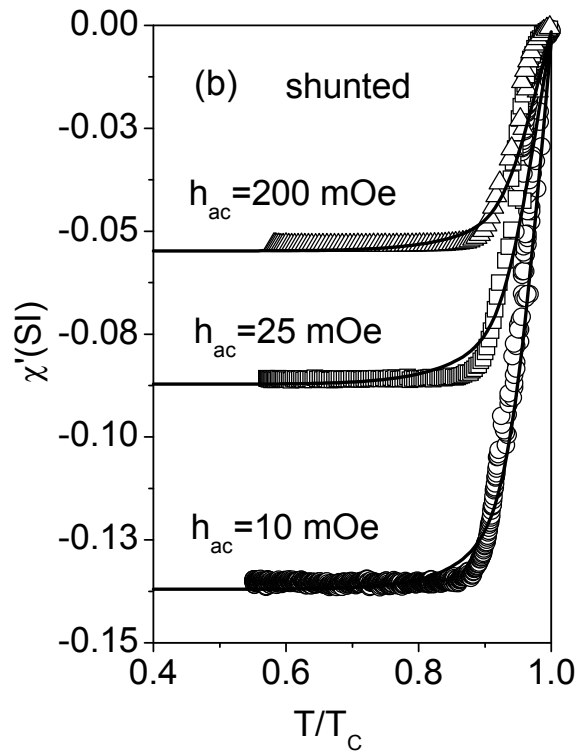

**Figure 7** - Experimental results for $\chi'(T, h_{ac}, H_{dc})$: (a) unshunted 2D-JJA for $h_{ac} = 10$ and 100 mOe; (b) shunted 2D-JJA for $h_{ac} = 10$, 25, and 200 mOe. In all these experiments $H_{dc} = 0$. Solid lines are the best fits (see text).

Since our shunted and unshunted samples have the same value of $\beta_L$ and different values of $\beta_C$, it is possible to verify the dependence of the reentrance effect on the value of the Stewart-McCumber parameter. For the unshunted 2D-JJA (Fig. 7a) we have found that for an AC field lower than 50 mOe (when the array is in the Meissner-like state) the behavior of $\chi'(T)$ is quite similar to homogeneous superconducting samples, while for $h_{ac} > 50$ mOe (when the array is in the mixed-like state with practically homogeneous flux distribution) these samples exhibit a clear reentrant behavior of susceptibility[44]. At the same time, the identical experiments performed on the shunted samples produced no evidence of any reentrance for all values of $h_{ac}$ (see Fig. 7b). It is important to point out that the analysis of the experimentally obtained imaginary component of susceptibility $\chi''(T)$ shows that for the highest AC magnetic field amplitudes (of about 200 mOe) dissipation remains small. Namely, for typical values of the AC amplitude, $h_{ac} = 100$ mOe (which corresponds to about 10 vortices per unit cell) the imaginary component is about 15 times smaller than its real counterpart. Hence contribution from the dissipation of vortices to the observed phenomena can be safely neglected.

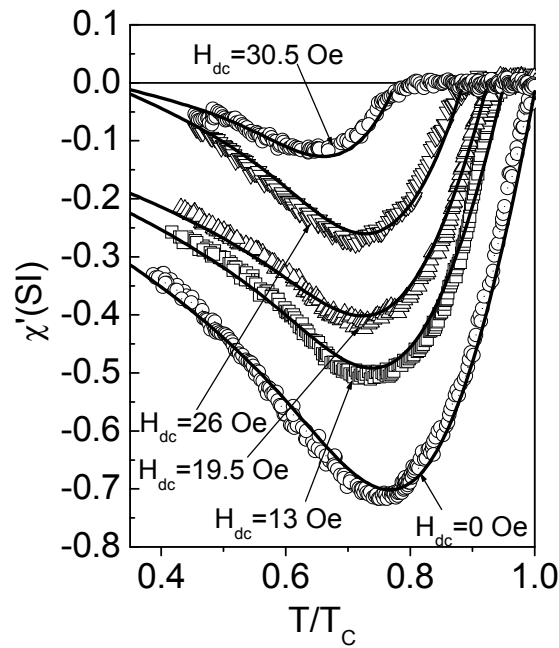

**Figure 8** - Experimental results for $\chi'(T, h_{ac}, H_{dc})$ for unshunted 2D-JJA for $H_{dc}$ = 0, 13, 19.5, 26, and 30.5 Oe. In all these experiments $h_{ac} = 100$ mOe. Solid lines are the best fits (see text).

To further study this unexpected behavior we have also performed experiments where we measure $\chi'(T)$ for different values of $H_{dc}$ keeping the value of $h_{ac}$ constant. The influence of DC fields on reentrance in unshunted samples is shown in Fig. 8. On the other hand, the shunted samples still show no signs of reentrance, following a familiar pattern of field-induced gradual diminishing of superconducting phase (very similar to a zero DC field flat-like behavior seen in Fig.7b).

To understand the influence of DC field on reentrance observed in unshunted arrays, it is important to emphasize that for our sample geometry this parallel field suppresses the critical current $I_C$ of each junction without introducing any detectable flux into the plaquettes of the array. Thus, a parallel DC magnetic field allows us to vary $I_C$ independently from temperature and/or applied perpendicular AC field. The measurements show (see Fig. 8) that the position of the reentrance is tuned by $H_{dc}$.

We also observe that the value of temperature $T_{min}$ (at which $\chi'(T)$ has a minimum) first shifts towards lower temperatures as we raise $H_{dc}$ (for small DC fields) and then bounces back (for higher values of $H_{dc}$). This non-monotonic behavior is consistent with the weakening of $I_C$ and corresponds to Fraunhofer-like dependence of the Josephson junction critical current on DC magnetic field applied in the plane of the junction. We measured $I_C$ from transport current-voltage characteristics, at different values of $H_{dc}$ at T = 4.2 K and found that $\chi'(T = 4.2K)$, obtained from the isotherm T = 4.2 K (similar to that given in Fig. 8), shows the same Fraunhofer-like dependence on $H_{dc}$ as the critical current $I_C(H_{dc})$ of the junctions forming the array (see Fig. 9). This gives further proof that only the junction critical current is varied in this experiment. This also indicates that the screening currents at low temperature (i.e., in the reentrant region) are proportional to the critical currents of the junctions. In addition, this shows an alternative way to obtain $I_C(H_{dc})$ dependence in big arrays. And finally, a sharp Fraunhofer-like pattern observed in both arrays clearly reflects a rather strong coherence (with negligible distribution of critical currents and sizes of the individual junctions) which is based on highly correlated response of *all* single junctions forming the arrays, thus proving their high quality. Such a unique behavior of Josephson junctions in our samples provides a necessary justification for suggested theoretical interpretation of the obtained experimental results. Namely, based on the above-mentioned properties of our arrays, we have found that practically all the experimental results can be explained by analyzing the dynamics of just a single unit cell in the array.

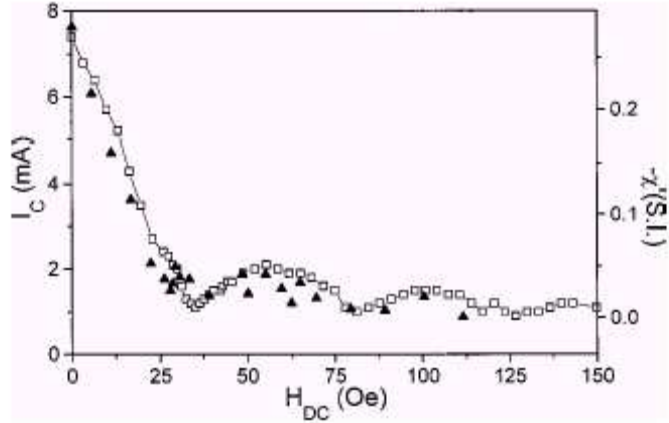

**Figure 9** - The critical current $I_C$ (open squares) and the real part of AC susceptibility $\chi'$ (solid triangles) as a function of DC field $H_{dc}$ for T=4.2K (from Ref.44).

To understand the different behavior of the AC susceptibility observed in shunted and unshunted 2D-JJAs, in principle one would need to analyze in detail the flux dynamics in these arrays. However, as we have previously reported[44], because of the well-defined periodic structure of our arrays (with no visible distribution of junction sizes and critical currents), it is reasonable to expect that the experimental results obtained from the magnetic properties of our 2D-JJAs can be quite satisfactory explained by analyzing the dynamics of a single unit cell (plaquette) of the array. An excellent agreement between a single-loop approximation and the observed behavior (seen through the data fits) justifies *a posteriori* our assumption. It is important to mention that the idea to use a single unit cell to qualitatively understand PME was first suggested by Auletta *et al.*[55]. They simulated the field-cooled DC magnetic susceptibility of a single-junction loop and found a paramagnetic signal at low values of external magnetic field.

In our calculations and numerical simulations, the unit cell is a loop containing four identical Josephson junctions and the measurements correspond to the zero-field cooling (ZFC) AC magnetic susceptibility. We consider the junctions of the single unit cell as having capacitance $C_j$, quasi-particle resistance $R_j$ and critical current $I_C$. As shown in previous sections, here we have also used this simple four-junctions model to study the magnetic behavior of our 2D-JJA by calculating the AC complex magnetic susceptibility $\chi = \chi'+i\chi''$ as a function of T, $\beta_L$ and $\beta_C$. Specifically, shunted samples are identified through low values of the McCumber parameter $\beta_C \approx 1$ while high values $\beta_C \gg 1$ indicate an unshunted 2D-JJA.

If we apply an AC external field $B_{ac}(t) = \mu_0 h_{ac} \cos(\omega t)$ normally to the 2D-JJA and a DC field $B_{dc} = \mu_0 H_{dc}$ parallel to the array, then the total magnetic flux $\Phi(t)$ threading the four-junction superconducting loop is given by $\Phi(t) = \Phi_{ext}(t) + LI(t)$ where L is the loop inductance, $\Phi_{ext}(t) = SB_{ac}(t) + (ld)B_{dc}$ is the flux related to the applied magnetic field (with $l \times d$ being the size of the single junction area, and $S \approx a^2$ being the projected area of the loop), and the circulating current in the loop reads:

$$I(t) = I_C(T)\sin\phi_i(t) + \frac{\Phi_0}{2\pi R_j}\frac{d\phi_i}{dt} + \frac{C_j\Phi_0}{2\pi}\frac{d^2\phi_i}{dt^2} \quad (V.1)$$

Here $\phi_i(t)$ is the gauge-invariant superconducting phase difference across the $i^{th}$ junction, and $\Phi_0$ is the magnetic flux quantum.

Since the inductance of each loop is $L = \mu_0 a \approx 64$ pH, and the critical current of each junction is $I_C \approx 150\mu A$, for the mixed-state region (above 50 mOe) we can safely neglect the self-field effects because in this region $LI(t)$ is always smaller than $\Phi_{ext}(t)$. Besides, since the length l and the width w of each junction in our array is smaller than the Josephson penetration depth $\lambda_j = \sqrt{\Phi_0/2\pi\mu_0 dj_{c0}}$ (where $j_{c0}$ is the critical current density of the junction, and $d = 2\lambda_L + \xi$ is the size of the contact area with $\lambda_L(T)$ being the London penetration depth of the junction and $\xi$ an insulator thickness), namely $l \approx w \approx 5\mu m$ and $\lambda_j \approx 20\mu m$ (using $j_{c0} \approx 600 A/cm^2$ and $\lambda_L \approx 39nm$ for Nb at T = 4.2 K), we can adopt the small-junction approximation[50] for the gauge-invariant superconducting phase difference across the $i^{th}$ junction (for simplicity we assume as usual[44] that $\phi_1 = \phi_2 = \phi_3 = \phi_4 \equiv \phi_i$):

$$\phi_i(t) = \phi_0(H_{dc}) + \frac{2\pi B_{ac}(t)S}{\Phi_0} \qquad (V.2)$$

where $\phi_0(H_{dc}) = \phi_0(0) + 2\pi\mu_0 H_{dc} ld/\Phi_0$ with $\phi_0(0)$ being the initial phase difference.

To properly treat the magnetic properties of the system, let us introduce the following Hamiltonian:

$$H(t) = J\sum_{i=1}^{4}[1-\cos\phi_i(t)] + \frac{1}{2}LI(t)^2 \qquad (V.3)$$

which describes the tunneling (first term) and inductive (second term) contributions to the total energy of a single plaquette. Here, $J(T) = (\Phi_0/2\pi)I_C(T)$ is the Josephson coupling energy.

The real part of the complex AC susceptibility is defined as:

$$\chi'(T, h_{ac}, H_{dc}) = \frac{\partial M}{\partial h_{ac}} \qquad (V.4)$$

where:

$$M(T, h_{ac}, H_{dc}) = -\frac{1}{V}\left\langle\frac{\partial H}{\partial h_{ac}}\right\rangle \qquad (V.5)$$

is the net magnetization of the plaquette. Here V is the sample's volume, and <...> denotes the time averaging over the period $2\pi/\omega$, namely:

$$\langle A \rangle = \frac{1}{2\pi}\int_0^{2\pi} d(\omega t) A(t) \qquad (V.6)$$

Taking into account the well-known[56] analytical approximation of the BCS gap parameter (valid for all temperatures), $\Delta(T) = \Delta(0)\tanh(2.2\sqrt{(T_C - T)/T})$ for the explicit temperature dependence of the Josephson critical current:

$$I_C(T) = I_C(0)\left[\frac{\Delta(T)}{\Delta(0)}\right]\tanh\left[\frac{\Delta(T)}{2k_B T}\right] \qquad (V.7)$$

we successfully fitted all our data using the following set of parameters: $\phi_0(0) = \pi/2$ (which corresponds to the maximum Josephson current within a plaquette), $\beta_L(0) = \beta_C(0) = 32$ (for unshunted array) and $\beta_C(0) = 1.2$ (for shunted array). The corresponding fits are shown by solid lines in Figs.7 and 8 for the experimental values of AC and DC field amplitudes.

In the mixed-state region and for low enough frequencies (this assumption is well-satisfied because in our case $\omega \ll \omega_{LR}$ and $\omega \ll \omega_{LC}$ where $\omega_{LR} = R/L$ and $\omega_{LC} = 1/\sqrt{LC}$ are the two characteristic frequencies of the problem) from Eqs.(V.3)-(V.6) we obtain the following approximate analytical expression for the susceptibility of the plaquette:

$$\chi'(T, h_{ac}, H_{dc}) \approx -\chi_0(T)\left[\beta_L(T)f_1(b)\cos\left(\frac{2H_{dc}}{H_0}\right) + f_2(b)\sin\left(\frac{H_{dc}}{H_0}\right) - \beta_C^{-1}(T)\right] \quad (V.8)$$

where $\chi_0(T) = \pi S^2 I_C(T)/V\Phi_0$, $H_0 = \Phi_0/(2\pi\mu_0 dl) \approx 10$ Oe, $f_1(b) = J_0(2b) - J_2(2b)$, $f_2(b) = J_0(b) - bJ_1(b) - 3J_2(b) + bJ_3(b)$ with $b = 2\pi S\mu_0 h_{ac}/\Phi_0$ and $J_n(x)$ being the Bessel function of the $n^{th}$ order.

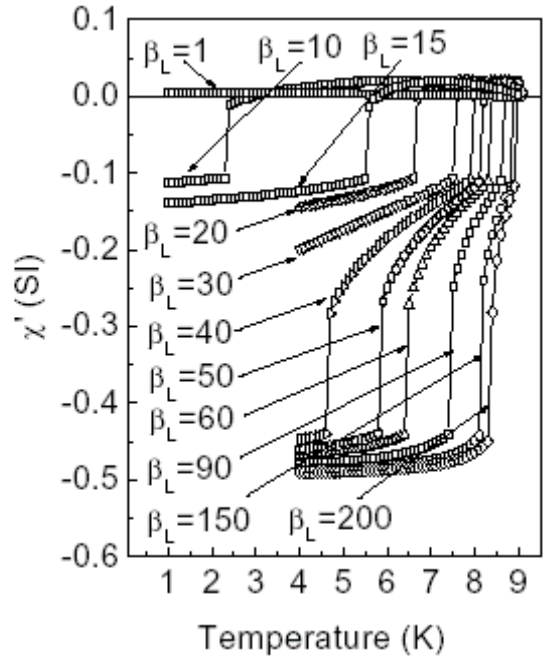

**Figure 10** - Numerical simulation results for $h_{ac} = 70$ mOe, $H_{dc} = 0$, $\beta_C(T = 4.2K) = 1$ and for different values of $\beta_L(T = 4.2K)$ based on Eqs.(V.4)-(V.7).

Notice also that the analysis of Eq.(V.8) reproduces the observed Fraunhofer-like behavior of the susceptibility in applied DC field (see Fig.9) and the above-mentioned fine tuning of the reentrance effect (see also Ref. 44). Indeed, according to Eq.(V.8) (and in agreement with the observations), for small DC fields the minimum temperature $T_{min}$ (indicating the beginning of the reentrant transition) varies with $H_{dc}$ as follows, $(T_C - T_{min})/T_C \approx H_{dc}/H_0$.

To further test our interpretation and verify the influence of the parameter $\beta_C$ on the reentrance, we have also performed extensive numerical simulations of the four-junction model previously described but without a simplifying assumption about the explicit form of the phase difference based on Eq.(V.2). More precisely, we obtained the temperature behavior of the susceptibility by solving the set of equations responsible for the flux dynamics within a single plaquette and based on Eq.(V.1) for the total current $I(t)$, the equation for the total flux $\Phi(t) = \Phi_{ext}(t) + LI(t)$ and the flux quantization condition for four junctions, namely $\phi_i(t) = (\pi/2)[n + (\Phi/\Phi_0)]$ where n is an integer. Both Euler and fourth-order Runge-Kutta integration methods provided the same numerical results. In Fig. 10 we show the real component of the simulated susceptibility $\chi(T)$ corresponding to the fixed value of $\beta_C(T = 4.2K) = 1$ (shunted samples) and different values of $\beta_L(T = 4.2K) = 1, 10, 15, 20, 30, 40, 50, 60, 90, 150$ and $200$. As expected, for this low value of $\beta_C$ reentrance is not observed for any values of $\beta_L$. On the other hand, Fig. 11 shows the real component of the simulated $\chi(T)$ but now using fixed value of $\beta_L(T = 4.2K) = 30$ and different values of $\beta_C(T = 4.2K) = 1, 2, 5, 10, 20, 30$ and $100$. This figure clearly shows that reentrance appears for values of $\beta_C > 20$. In both cases we used $h_{ac}$=70 mOe. We have also simulated the curve for shunted ($\beta_L = 30$, $\beta_C = 1$) and unshunted ($\beta_L = 30$, $\beta_C = 30$) samples for different values of $h_{ac}$ (see Fig. 12). In this case the values of the parameters $\beta_L$ and $\beta_C$ were chosen from our real 2D-JJA samples. Again, our simulations confirm that dynamic reentrance does not occur for low values of $\beta_C$, independently of the values of $\beta_L$ and $h_{ac}$.

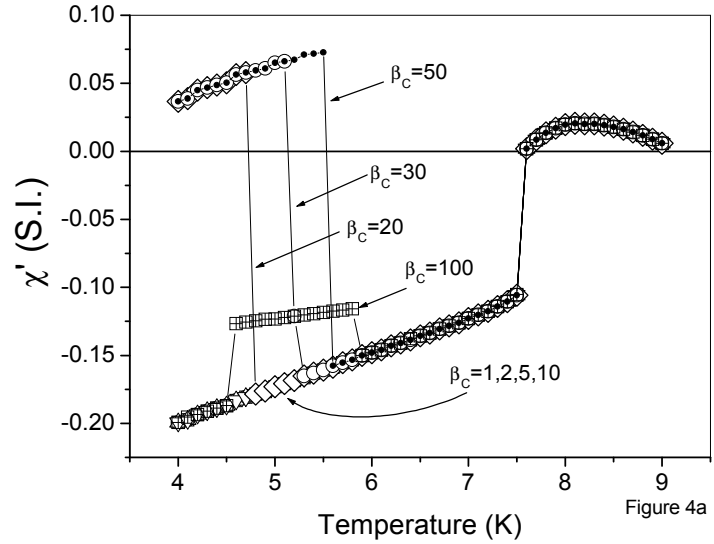

**Figure 11** - Numerical simulation results for $h_{ac} = 70$ mOe, $H_{dc} = 0$, $\beta_L(T = 4.2K) = 30$ and for different values of $\beta_C(T = 4.2K)$ based on Eqs.(V.4)-(V.7).

The following comment is in order regarding some irregularities ("jumps" and "steps") visibly seen in Figs.(10)-(12) around the transition regions from non-reentrant to reentrant behavior. It is important to emphasize that the above irregularities are just artifacts of the numerical simulations due to the conventional slow-converging real-time reiteration procedure[44]. They neither correspond to any experimentally observed behavior (within the accuracy of the measurements technique and data acquisition), nor they reflect any irregular features of the considered here theoretical model (which predicts a smooth temperature dependence seen through the data fits). As usual, to avoid this kind of artificial (non-physical) discontinuity, more powerful computers are needed.

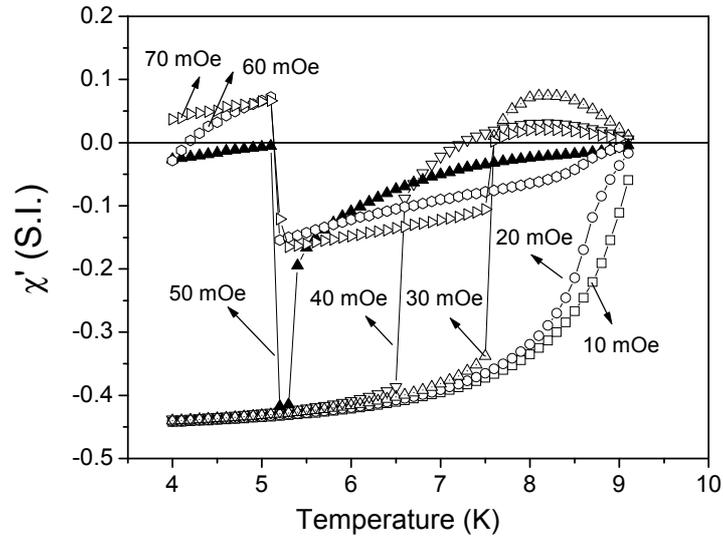

(a)

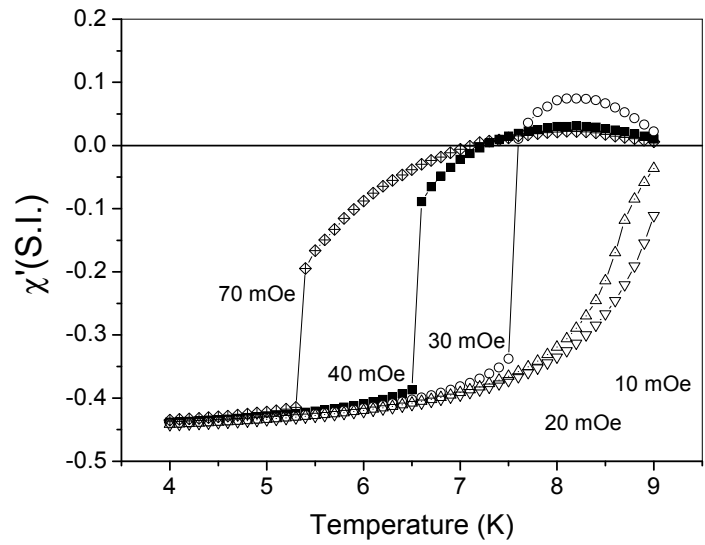

(b)

**Figure 12** - Curves of the simulated susceptibility ($H_{dc} = 0$ and for different values of $h_{ac}$) corresponding to (a) unshunted 2D-JJA with $\beta_L(T = 4.2K) = 30$ and $\beta_C(T = 4.2K) = 30$; (b) shunted 2D-JJA with $\beta_L(T = 4.2K) = 30$ and $\beta_C(T = 4.2K) = 1$.

Based on the above extensive numerical simulations, a resulting *phase diagram* $\beta_C$-$\beta_L$ (taken for T=1K, $h_{ac}$=70 mOe, and $H_{dc}$=0) is depicted in Fig. 13 which clearly demarcates the border between the reentrant (white area) and non-reentrant (shaded area) behavior in the arrays for different values of $\beta_L(T)$ and $\beta_C(T)$ parameters at given temperature. In other words, if $\beta_L$ and $\beta_C$ parameters of any realistic array have the values inside the white area, this array will exhibit a reentrant behavior.

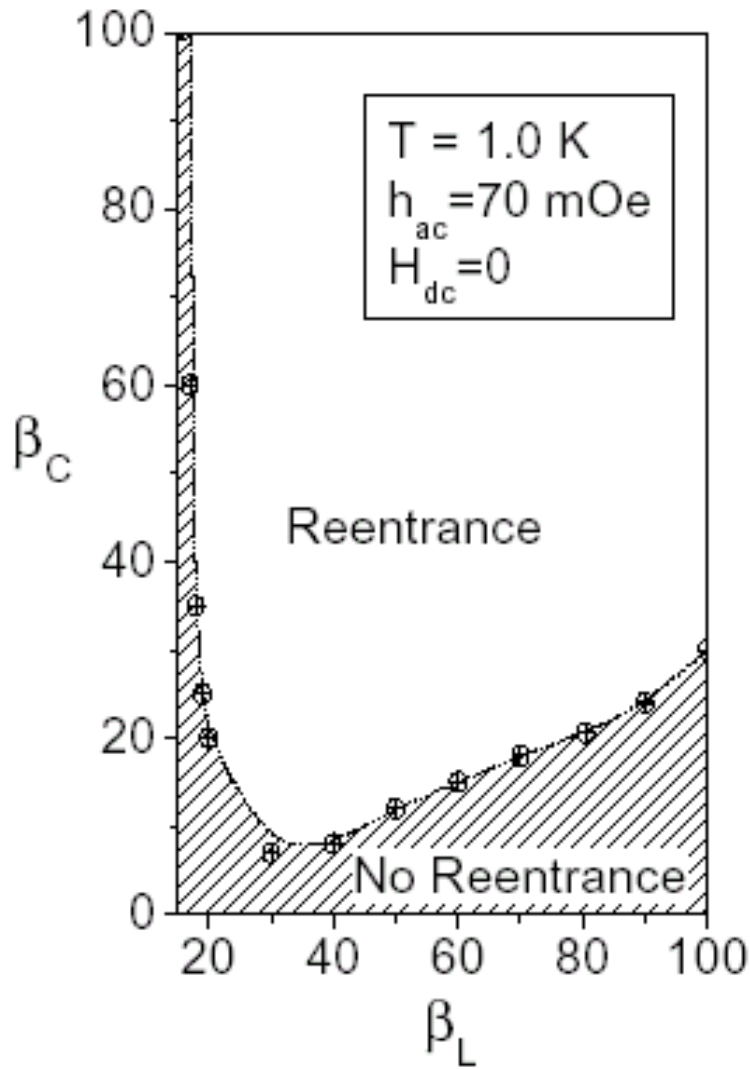

**Figure 13** - Numerically obtained *phase diagram* (taken for $T=1K$, $h_{ac} = 70$ mOe, and $H_{dc}=0$) which shows the border between the reentrant (white area) and non-reentrant (shaded area) behavior in the arrays for different values of $\beta_L$ and $\beta_C$ parameters.

It is instructive to mention that a hyperbolic-like character of $\beta_L$ vs. $\beta_C$ law (seen in Fig. 13) is virtually present in the approximate analytical expression for the susceptibility of the plaquette given by Eq.(V.8) (notice however that this expression can not be used to produce any quantitative prediction because the neglected in Eq.(V.8) frequency-related terms depend on $\beta_L$ and $\beta_C$ parameters as well). A qualitative behavior of the envelope of the *phase diagram* (depicted in Fig. 13) with DC magnetic field $H_{dc}$ (for T=1 K and $h_{ac}$=70 mOe), obtained using Eq.(V.8), is shown in Fig. 14.

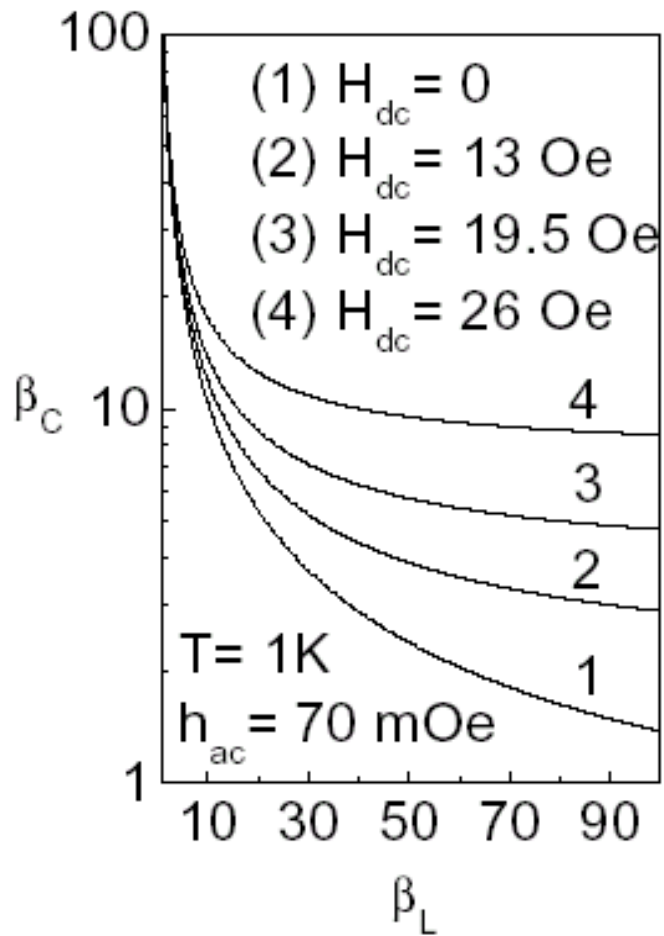

**Figure 14** - A qualitative behavior of the envelope of the *phase diagram* (shown in previous figure) with DC magnetic field $H_{dc}$ (for $T = 1K$ and $h_{ac} = 70mOe$) obtained from Eq.(V.8).

And finally, to understand how small values of $\beta_C$ parameter affect the flux dynamics in shunted arrays, we have analyzed the $\Phi_{tot} - \Phi_{ext}$ diagram. Similarly to those results previously obtained from unshunted samples[44], for a shunted sample at fixed temperature this curve is also very hysteretic (see Fig. 15). In both cases, $\Phi_{tot}$ vs. $\Phi_{ext}$ shows multiple branches intersecting the line $\Phi_{tot} = 0$ which corresponds to diamagnetic states. For all the other branches, the intersection with the line $\Phi_{tot} = \Phi_{ext}$ corresponds to the boundary between diamagnetic states (negative values of $\chi'$) and paramagnetic states (positive values of $\chi'$). As we have reported before[44], for unshunted 2D-JJA at temperatures below 7.6 K the appearance of the first and third branches adds a paramagnetic contribution to the average value of $\chi'$. When $\beta_C$ is small (shunted arrays), the analysis of these curves shows that there is no reentrance at low temperatures because in this case the second branch appears to be energetically stable, giving an extra diamagnetic contribution which overwhelms the paramagnetic contribution from subsequent branches. In other words, for low enough values of $\beta_C$ (when the samples are ZFC and then measured at small values of the magnetic field), most of the loops will be in the diamagnetic states, and no paramagnetic response is registered. As a result, the flux quanta cannot get trapped into the loops even by the following field-cooling process in small values of the magnetic field. In this case the superconducting phases and the junctions will have the same diamagnetic response and the resulting measured value of the magnetic susceptibility will be negative (i.e., diamagnetic) as well. On the other hand, when $\beta_C$ is large enough (unshunted arrays), the second branch becomes energetically unstable, and the average response of the sample at low temperatures is paramagnetic (Cf. Fig. 7 from Ref. 44).

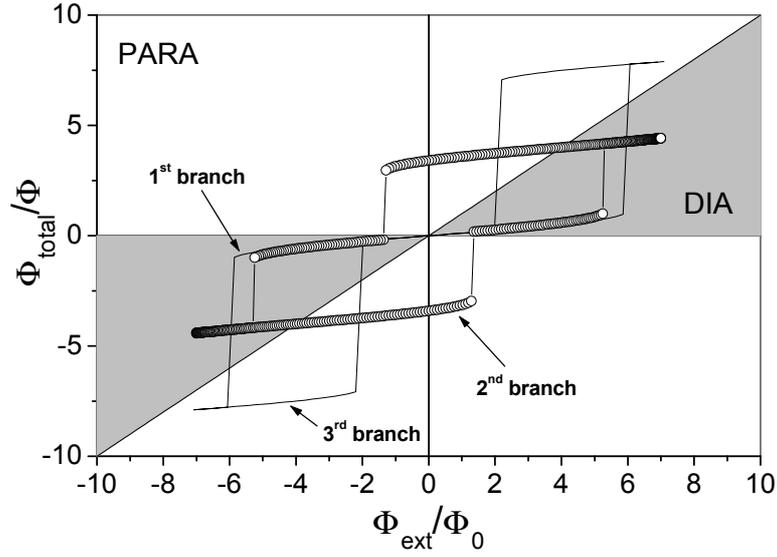

**Figure 15** - Numerical simulation results, based on Eqs.(V.4)-(V.7), showing $\Phi_{tot}$ vs. $\Phi_{ext}$ for shunted 2D-JJA with $\beta_L(T=4.2K)=30$ and $\beta_C(T=4.2K)=1$.

In summary, in this section we have shown that our experimental and theoretical results demonstrate that the reentrance phenomenon (and concomitant PME) in artificially prepared Josephson Junction Arrays is related to the damping effects associated with the Stewart-McCumber parameter $\beta_C$. Namely, reentrant behavior of AC susceptibility takes place in the underdamped (unshunted) array (with large enough value of $\beta_C$) and totally disappears in overdamped (shunted) arrays.

# VI. Manifestation of novel geometric effects in temperature behavior of AC magnetic response.

Many unusual and still not completely understood magnetic properties of 2D-JJAs continue to attract attention of both theoreticians and experimentalists alike (for recent reviews on the subject see, e.g. Refs. [57-61] and further references therein). In particular, among the numerous spectacular phenomena recently discussed and observed in 2D-JJAs we would like to mention the dynamic temperature reentrance of AC susceptibility[57,58] (closely related to paramagnetic Meissner effect[59]) and avalanche-like magnetic field behavior of magnetization[60,61] (closely related to self-organized criticality (SOC)[62,63]). More specifically, using highly sensitive SQUID magnetometer, magnetic field jumps in the magnetization curves associated with the entry and exit of avalanches of tens and hundreds of fluxons were clearly seen in SIS-type arrays[61]. Besides, it was shown that the probability distribution of these processes is in good agreement with the SOC theory[63]. An avalanche character of flux motion was observed at temperatures at which the size of the fluxons did not exceed the size of the cell, that is, for discrete vortices. On the other hand, using a similar technique, magnetic flux avalanches were not observed in SNS-type proximity arrays[64] despite a sufficiently high value of the inductance L related critical parameter $\beta_L = 2\pi L I_C / \Phi_0$ needed to satisfy the observability conditions of SOC. Instead, the observed quasi-hydrodynamic flux motion in the array was explained by the considerable viscosity characterizing the vortex motion through the Josephson junctions.

In this section of the present review article, we show experimental evidence for manifestation of novel geometric effects in magnetic response of high-quality ordered 2D-JJA. By improving resolution of home-made mutual-inductance measurements technique described in the beginning of this article, a pronounced step-like structure (with the number of steps n = 4 for all AC fields) has been observed in the temperature dependence of AC susceptibility in artificially prepared two-dimensional Josephson Junction Arrays (2D-JJA) of unshunted Nb-AlO$_x$-Nb junctions with $\beta_L(4.2K) = 30$. Using a single-plaquette approximation of the overdamped 2D-JJA model, we were able to successfully fit our data assuming that steps are related to the geometric properties of the plaquette. The number of steps n corresponds to the number of flux quanta that can be screened by the maximum critical current of the junctions. The steps are predicted to manifest themselves in arrays with the inductance related parameter $\beta_L$ matching a "quantization" condition $\beta_L(0) = 2\pi(n+1)$.

To measure the complex AC susceptibility in our arrays with high precision, we used a home-made susceptometer based on the so-called screening method in the reflection configuration as described in the previous sections[65-67]. Measurements were performed as a function of the temperature T (for 1.5 K < T < 15 K), and the amplitude of the excitation field $h_{ac}$ (for 1 mOe < $h_{ac}$ < 10 Oe) normal to the plane of the array. The frequency of AC field in the experiments reported here was fixed at 20 kHz. The used in the present study unshunted 2D-JJAs are formed by loops of niobium islands (with $T_C$ = 9.25 K) linked through Nb-AlO$_x$-Nb Josephson junctions and consist of 100×150 tunnel junctions described in previous sections.

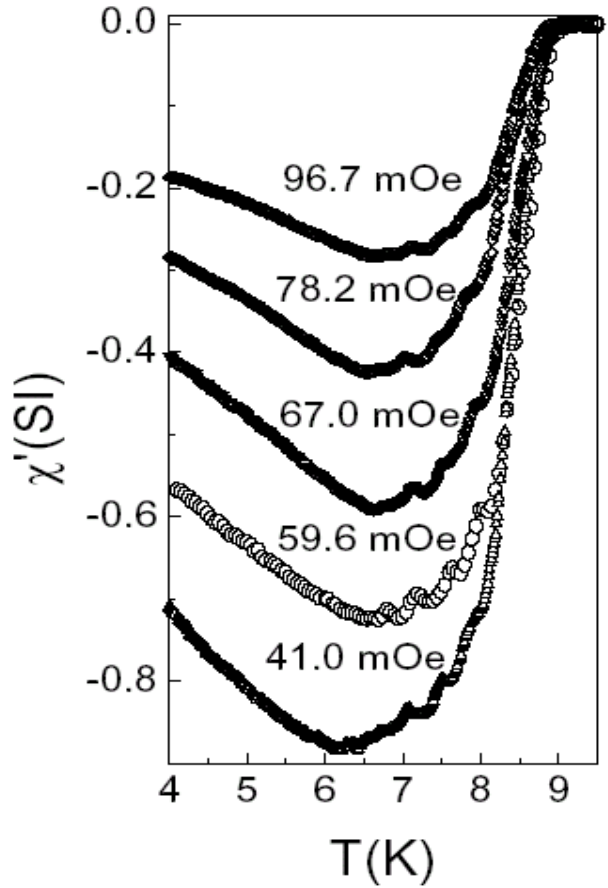

**Figure 16** - Experimental results for temperature dependence of the real part of AC susceptibility $\chi'(T, h_{ac})$ for different AC field amplitudes $h_{ac}$ = 41.0, 59.6, 67.0, 78.2 and 96.7 mOe.

It is important to recall that the magnetic field behavior of the critical current of the array (taken at T=4.2 K) on DC magnetic field $H_{dc}$ (parallel to the plane of the sample) exhibited a sharp Fraunhofer-like pattern characteristic of a single-junction response, thus proving a rather strong coherence within arrays (with negligible distribution of critical currents and sizes of the individual junctions) and hence the high quality of our sample.

The observed temperature dependence of the real part of AC susceptibility for different AC fields is shown in Fig. 16. A pronounced step-like structure is clearly seen at higher temperatures. The number of steps n does not depend on AC field amplitude and is equal to n = 4. As expected[58,67,68], for $h_{ac}$ > 40 mOe (when the array is in the mixed-like state with practically homogeneous flux distribution) the steps are accompanied by the previously observed reentrant behavior with $\chi'(T, h_{ac})$ starting to increase at low temperatures.

To understand the step-like behavior of the AC susceptibility observed in unshunted 2D-JJAs, in principle one would need to analyze in detail the flux dynamics in these arrays. However, as we have previously reported[58,67,68], because of the well-defined periodic structure of our arrays with no visible distribution of junction sizes and critical currents, it is quite reasonable to assume that the experimental results obtained from the magnetic properties of our 2D-JJAs could be understood by analyzing the dynamics of just a single unit cell (plaquette) of the array. As we shall see, theoretical interpretation of the presented here experimental results based on single-loop approximation, is in excellent agreement with the observed behavior. In our analytical calculations, the unit cell is the loop containing four identical Josephson junctions described in previous sections, and the measurements correspond to the zero-field cooling AC magnetic susceptibility. If we apply an AC external field $H_{ac}(t) = h_{ac}\cos\omega t$ normally to the 2D-JJA, then the total magnetic flux $\Phi(t)$ threading the four-junction superconducting loop is given again by $\Phi(t) = \Phi_{ext}(t) + LI(t)$ where L is the loop inductance, $\Phi_{ext}(t) = SH_{ac}(t)$ is the flux related to the applied magnetic field (with $S \approx a^2$ being the projected area of the loop), and the circulating current in the loop reads $I(t) = I_C(T)\sin\phi(t)$. Here $\phi(t)$ is the gauge-invariant superconducting phase difference across the i$^{th}$ junction. As is well-known, in the case of four junctions, the flux quantization condition reads[67,69]

$$\phi = \frac{\pi}{2}\left(n + \frac{\Phi}{\Phi_0}\right) \qquad (VI.1)$$

where *n* is an integer, and, for simplicity, we assume as usual that[58,67] $\phi_1 = \phi_2 = \phi_3 = \phi_4 \equiv \phi$.

To properly treat the magnetic properties of the system, let us introduce the following Hamiltonian

$$H(t) = J(T)[1 - \cos\phi(t)] + \frac{1}{2}LI^2(t) \qquad (VI.2)$$

which describes the tunneling (first term) and inductive (second term) contributions to the total energy of a single plaquette. Here, $J(T) = (\Phi_0/2\pi)I_C(T)$ is the Josephson coupling energy.

Since the origin of reentrant behavior in our unshunted arrays has been discussed in much detail earlier[58,67,68] (see also the previous section of this Chapter), in what follows we concentrate only on interpretation of the observed here step-like structure of $\chi'(T, h_{ac})$. First of all, we notice that the number of observed steps n (in our case n = 4) clearly hints at a possible connection between the observed here phenomenon and flux quantization condition within a single four-junction plaquette. Indeed, the circulating in the loop current $I(t) = I_C(T)\sin\phi(t)$ passes through its maximum value whenever $\phi(t)$ reaches the value of $(\pi/2)(2n+1)$ with n = 0,1,2... As a result, the maximum number of fluxons threading a single plaquette (see Eq. (VI.1)) over the period $2\pi/\omega$ becomes equal to $<\Phi(t)> = (n+1)\Phi_0$. In turn, the latter equation is equivalent to the following condition $\beta_L(T) = 2\pi(n+1)$. Since this formula is valid for any temperature, we can rewrite it as a geometrical "quantization" condition $\beta_L(0) = 2\pi(n+1)$. Recall that in the present experiment, our array has $\beta_L(0) = 31.6$ (extrapolated from its experimental value $\beta_L(4.2K) = 30$) which is a perfect match for the above "quantization" condition predicting n = 4 for the number of steps in a single plaquette, in excellent agreement with the observations.

Based on the above discussion, we conclude that in order to reproduce the observed temperature steps in the behavior of AC susceptibility, we need a particular solution to Eq.(VI.1) for the phase difference in the form of $\phi_n(t) = (\pi/2)(2n+1) + \delta\phi(t)$ assuming $\delta\phi(t) \ll 1$. After substituting this Ansatz into Eq.(VI.1), we find that $\phi_n(t) \approx (\pi/2)n + (1/4)\beta_L(T) + (1/4)f\cos(\omega t)$ where $f = 2\pi S h_{ac}/\Phi_0$ is the AC field related frustration parameter. Using this effective phase difference, we can calculate the AC response of a single plaquette. Namely, the real part of susceptibility reads

$$\chi'(T, h_{ac}) = \frac{1}{\pi}\int_0^\pi d(\omega t)\cos(\omega t)\chi_n(t) \qquad (VI.3)$$

where

$$\chi_n(t) = -\frac{1}{V}\left[\frac{\partial^2 H}{\partial h_{ac}^2}\right]_{\phi=\phi_n(t)} \qquad (VI.4)$$

Here V is the sample's volume.

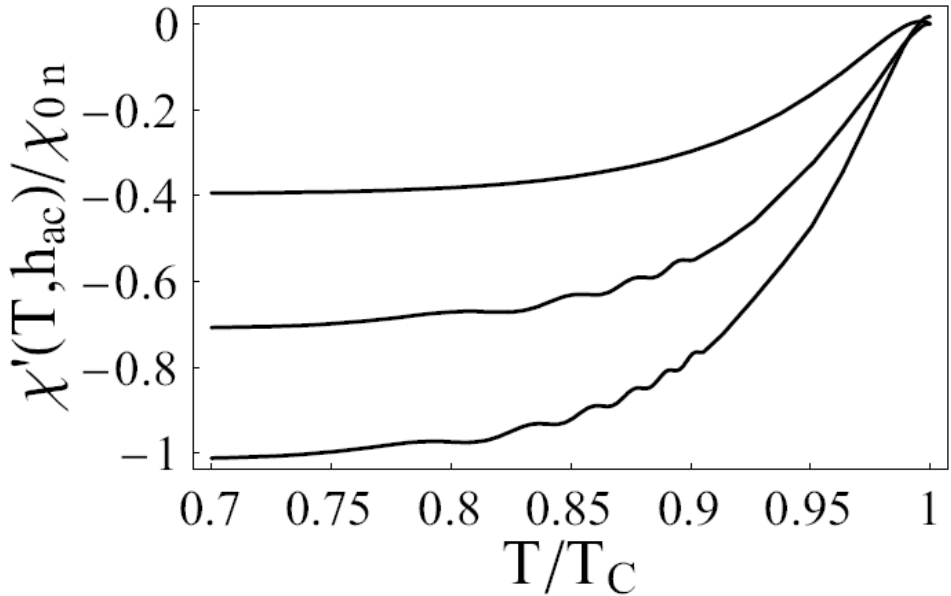

**Figure 17** - Theoretically predicted dependence of the normalized susceptibility on reduced temperature according to Eqs.(VI.3)-(VI.5) for f=0.5 and for "quantized" values of $\beta_L(0) = 2\pi(n+1)$ (from top to bottom): n=0, 3 and 5.

For the explicit temperature dependence of $\beta_L(T) = 2\pi L I_C(T)/\Phi_0$ we use again the well-known[70,71] analytical approximation of the BCS gap parameter (valid for all temperatures), $\Delta(T) = \Delta(0)\tanh(2.2\sqrt{(T_C - T)/T})$ which governs the temperature dependence of the Josephson critical current:

$$I_C(T) = I_C(0)\left[\frac{\Delta(T)}{\Delta(0)}\right]\tanh\left[\frac{\Delta(T)}{2k_BT}\right] \qquad (VI.5)$$

Fig. 17 depicts the predicted by Eqs.(VI.3)-(VI.5) dependence of the AC susceptibility on reduced temperature for f=0.5 and for different "quantized" values of $\beta_L(0) = 2\pi(n+1)$. Notice the clear appearance of three and five steps for n = 3 and n = 5, respectively (as expected, n = 0 corresponds to a smooth temperature behavior without steps).

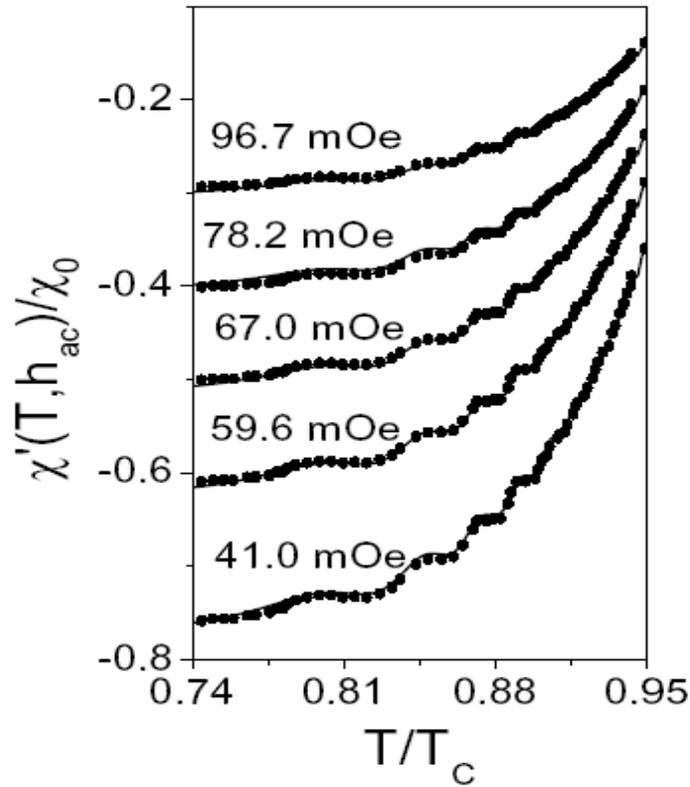

**Figure 18** - Fits (solid lines) of the experimental data for $h_{ac}$ = 41.0, 59.6, 67.0, 78.2, and 96.7 mOe according to Eqs.(VI.3)-(VI.5) with $\beta_L(0) = 10\pi$.

In Fig. 18 we present fits (shown by solid lines) of the observed temperature dependence of the normalized susceptibility $\chi'(T, h_{ac})/\chi_0$ for different magnetic fields $h_{ac}$ according to Eqs.(VI.3)-(VI.5) using $\beta_L(0) = 10\pi$. As is seen, our simplified model based on a single-plaquette approximation demonstrates an excellent agreement with the observations.

In conclusion, in this section we have shown a step-like structure (accompanied by previously seen low-temperature reentrance phenomenon) which has been observed for the first time in the temperature dependence of AC susceptibility in artificially prepared two-dimensional Josephson Junction Arrays of unshunted Nb-AlO$_x$-Nb junctions. The steps are shown to occur in arrays with the inductance related parameter $\beta_L(T)$ matching the "quantization" condition $\beta_L(0) = 2\pi(n+1)$ where n is the number of steps.

## VII. Summary

To summarize, in this review article we reported on three phenomena related to the magnetic properties of 2D-JJA: (a) the influence of non-uniform critical current density profile on magnetic field behavior of AC susceptibility; (b) the origin of dynamic reentrance and the role of the Stewart-McCumber parameter, $\beta_C$, in observability of this phenomenon, and (c) the manifestation of novel geometric effects in temperature behavior of AC magnetic response.

We have found clear experimental evidence for the influence of the junction non-uniformity on magnetic field penetration into the periodic 2D array of unshunted Josephson junctions. By using the well-known AC magnetic susceptibility technique, we have shown that in the mixed-state regime the AC field behavior of the artificially prepared array is reasonably well fitted by the single-plaquette approximation of the over-damped model of 2D-JJA assuming inhomogeneous (Lorentz-like) critical current distribution within a single junction.

On the other hand, our experimental and theoretical results have demonstrated that the reentrance of AC susceptibility (and concomitant PME) in artificially prepared Josephson Junction Arrays takes place in the underdamped (unshunted) array (with large enough value of the Stewart-McCumber parameter $\beta_C$) and totally disappears in over-damped (shunted) arrays.

Finally, we have shown a step-like structure (accompanied by previously seen low-temperature reentrance phenomenon) which has been observed for the first time in the temperature dependence of AC susceptibility in artificially prepared two-dimensional Josephson Junction Arrays of unshunted Nb-AlO$_x$-Nb junctions. The steps are shown to occur in arrays with the inductance related parameter $\beta_L(T)$ matching the "quantization" condition $\beta_L(0) = 2\pi(n+1)$ where n is the number of steps.

**Acknowledgements**

Very useful discussions with P. Barbara, C.J. Lobb, A. Sanchez and R.S. Newrock are highly appreciated. We thank W. Maluf for his help in running some of the experiments. We gratefully acknowledge financial support from Brazilian Agency FAPESP under grant 2003/00296-5.